\documentclass[12pt,preprint]{aastex}

\begin{document}

\title
{The Outer Galactic Halo As Probed By RR Lyr Stars \\ From the Palomar Transient Facility 
+ Keck\altaffilmark{1}}

\author
{Judith Cohen\altaffilmark{2}, Branimir Sesar\altaffilmark{3}, 
Sophianna Banholzer\altaffilmark{2}, the PTF Collaboration\altaffilmark{2} }

\altaffiltext{1}{Based in part on observations obtained at the
W.M. Keck Observatory, which is operated jointly by the California
Institute of Technology, the University of California, and the
National Aeronautics and Space Administration.}

\altaffiltext{2}{California Institute of Technology, Palomar Observatory, 
Mail Code 249-17, Pasadena, Ca., 91125, jlc@astro.caltech.edu}

\altaffiltext{3}{Max Planck Institute for Astronomy, Konigst{\"u}hl 17, D-69117,
Heidelberg, Germany, bsesar@mpia.de}

\begin{abstract}
We present initial results from our study of the outer halo of the Milky Way using a large 
sample of RR Lyr($ab$) 
variables datamined from the archives of the Palomar Transient Facility.  Of the 
464 RR Lyr in our sample with distances exceeding 50~kpc, 62 have been observed spectroscopically at the Keck Observatory.  $v_r$ and $\sigma(v_r)$ 
are given as a function of distance between 50 and 110~kpc, and
a very preliminary rather low total mass for the Milky Way out to 110~kpc of 
$\sim{7{\pm{1.5}}\times}10^{11}~M_{\odot}$ is derived
from our data.

\end{abstract}

\keywords{The Galaxy, Galaxy:halo, stars:variables:RR Lyrae}

\section{Introduction}

We present initial results from our study of the outer halo of the Milky Way (MW)
using a large sample of RR Lyr($ab$) variables datamined for the archives of the 
Palomar Transient Facility (PTF) (Law et al, 2009, Rau et al 2009)
(P.I. S.~R. Kulkarni of Caltech). 
RR Lyr are old low-mass pulsating stars with distinctive light curves, 
amplitudes at $V$ of $\sim 1$~mag, and periods of 
$\sim$0.5~days.
These characteristics makes them fairly easy to distinguish 
in a wide field, multi-epoch optical imaging survey of moderate duration if the survey cadence is suitable.  
Their most desirable
characteristic is that they can be used as standard candles.  Accurate luminosities, which have
only a small metallicity dependence, can be inferred directly from the light curves, and these
stars, with $M_V \sim +0.6$~mag, are fairly luminous and hence can be detected at large distances.

Using machine learning techniques we have isolated 
a sample of 464 RR Lyr from the PTF.  These were found by searching for
PTF fields which had enough epochs of observation (25 minimum) in the R filter.
A total of
roughly 10,000~deg$^2$ on the sky met this requirement and was searched.  
The regions of known outer halo objects (i.e. dwarf galaxies,
globular clusters, and known halo streams) were excluded.
The search criterion was variability, and the sample was refined by requiring
the derived period and amplitude to have values appropriate for RR Lyr.
Since no observing time was assigned for this purpose, we are effectively
datamining the PTF archive, and our sample lies in  
random pencil beams through the halo, each a PTF tile (area on sky: 7.6~deg$^2$).

Our sample begins at a brightness corresponding to a distance of 50~kpc,
and extends out to $\sim$110~kpc, after which RR Lyr are too faint 
to pick out with PTF data.  The selection procedure was carried out in 2012;
at present the PTF+iPTF variable star database contains photometry through
Dec 2014, i.e. two more years of data, none of which was used to select the
current sample.
A more detailed description of the search method is given in Sesar et al (2014).
Given this is a variability search, the only contaminants with colors similar
to those of RR Lyr are quasars (QSOs).

Fig.~1 (top) shows the light curve of one of the brighter RR Lyr
in our sample ($r = 55$~kpc) as well as that of one of the most distant RR Lyr 
found to date ($r = 102$~kpc) (bottom).

\begin{figure}
\epsscale{0.6}
\plotone{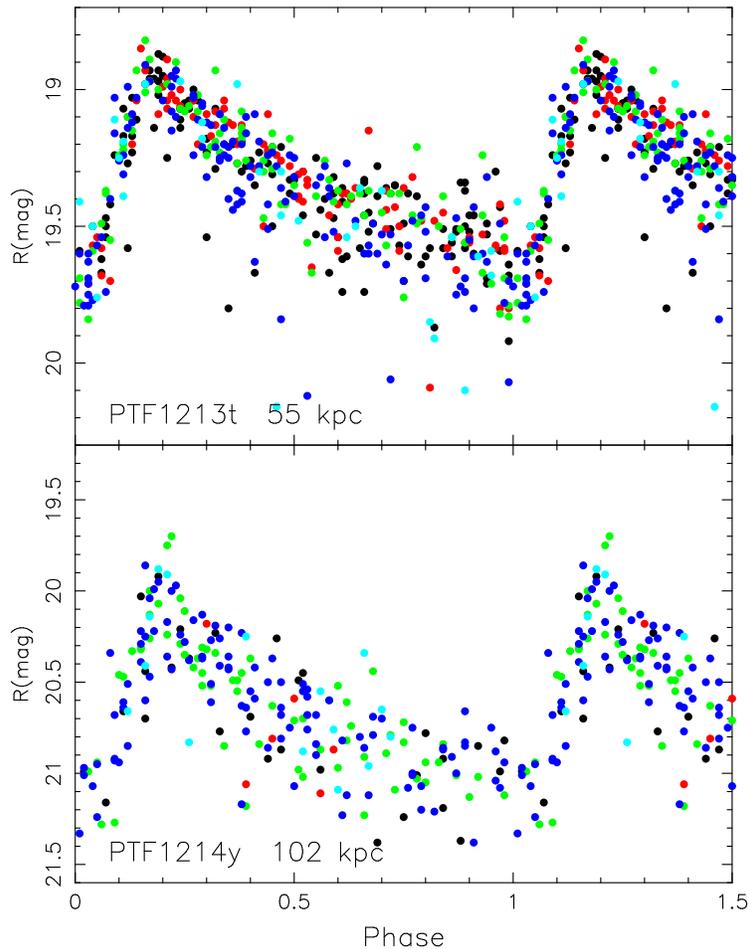}
\caption[]{Examples of phased PTF light curves: {\bf{Top:}} A RR Lyr at a distance of 55 kpc
with 468  measurements with the R filter in the PTF database.
{\bf{Bottom:}} A RR Lyr at a distance of 102 kpc with 219 R measurements in the PTF database. 
Note that prior to the addition of 2 years of data to the PTF variable star database
in early 2015,
this star had only 61 detections at R in the PTF database.
The color scheme 
denotes the year of observation, with the first day set to the first night
the object was observed by the PTF.  Black = year 1, red = year 2, green = year 3, blue = year 4,
etc using the PGPLOT color routine.  The sample selection was carried out using only the first
two years of PTF data.
   \label{figure_lightcurve} }
\end{figure}

\clearpage

\section{Radial Velocity Measurements}

A spectroscopic campaign to obtain radial velocities for RR Lyr candidates
began at the Keck Observatory with 
the Deimos spectrograph (Faber et al 2003) in the spring of 2014.
RR Lyr are pulsating variable stars, hence their observed radial velocities 
need to be corrected for the motion of the atmosphere.  To measure the center-of-mass
velocity $v_r$ of RR Lyr stars, we use the Balmer H$\alpha$ line and
the method described by Sesar (2012).
Typical uncertainties in center-of-mass velocities, including both observational
and phase correction, are 15 to 20~km/s for a single Deimos spectrum.

Fig.~2 shows $v_r$ vs $r$ for the 62 RR Lyr with Keck/Deimos spectra.
We cut the sample
at 85~kpc, and assume an uncertainty in each individual measurement
of 20~km/s.  We find a mean $v_r$ for the inner 41 RR Lyr of $-1$~km/s with
$\sigma = 116$~km/s, while for the outer 21 stars, the mean is $+13$~km/s with 
$\sigma(v_r) = 91$~km/s.   The mean of $v_r$ falls close to 0, suggesting that
our  corrections to observed radial velocities are appropriate.

When 4 major outliers (2 high and 2 low) are deleted
from the inner sample, $\sigma$ falls to 90~km/s
for  the inner sample. Deleting three high outliers from the
outer sample leaves 18 R Lyr beyond 85~kpc, and $\sigma$ falls to 57~km/s,
a remarkably low dispersion.  
We are not yet certain whether
the outliers arise from substructures within the outer halo.

The velocity dispersion as a function of radius for the Milky Way halo is
shown in Fig.~3, where we compare our work to selected values
from the recent literature.  The small stars denote our entire sample,
cut at 85~kpc,
and the larger stars show our sample of outer halo RR Lyr cleaned of outliers;  
recall that the outliers comprise only about 10\% of our total sample.  
The star symbols are plotted at the median distance of the inner and 
of the outer sample.
The velocity dispersions found for the inner halo from recent studies
are also shown:  from SDSS blue horizontal branch (BHB) stars
(Xue et al 2004), from SDSS K giants (Xue et al 2014) as well as
a sample of halo high velocity stars from
Brown et al (2010).  Most of these other samples are confined to distances
less than 80 kpc.  The only attempt to reach the distances probed by the
outer part of our sample is that of Deason et al (2012).  
Our results agree  quite well with
the value they published of $\sigma(v_r)$ of 50 to 60~km/s 
at distances of
$\sim$110~kpc, rising to $\sim$90~km/s at distances of 70~kpc.
We agree reasonably well with the result of Brown et al (2010) for $\sigma(v_r)$
as a function of $r$ from the MMT high velocity star survey, extending out to
75~kpc, when their non-parametric method to eliminate outliers is used
(see their Fig.~6).

\clearpage

\begin{figure}
\epsscale{0.4}
\plotone{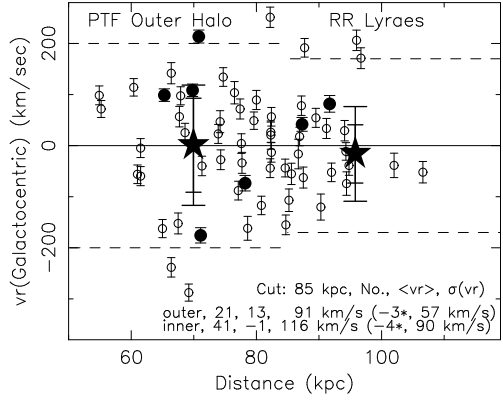} 
 \caption{Radial velocities as a function of heliocentric distances are shown
for our sample of 62 RR Lyr selected from the PTF with Keck/Deimos moderate
resolution spectra.  The borders of the regions (both high and low $v_r$) considered outliers
are indicated by the dashed horizontal lines.
The text at the bottom right gives the number of RR Lyr, mean and $\sigma$
(with 20~km/s removed in quadrature for observational uncertainties) for the inner
and for the outer sample, cut at 85~kpc, as well as these with the largest
outliers removed (3 stars for the outer region, and 4 for the closer sample).
Filled circles denote stars with two Deimos spectra, open circles have only
one spectrum.  One $\sigma$ error bars are shown for each RR Lyr.  
The two large stars denote the means for the inner and outer sample with the outliers removed;
their two error bars correspond to $\sigma$ for the full sample, and that
for the cropped sample.
   \label{figure_vr} }
\end{figure}

\begin{figure}
\epsscale{0.4}
\plotone{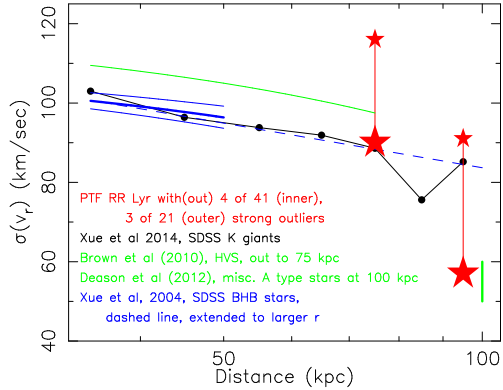}
 \caption{$\sigma(v_r)$ is shown as a function of log($r$) for
our inner and outer sample of RR Lyr (split at 85~kpc), with and without
eliminating the single strong outliers.  Values 
from \cite{xue08}, whose extrapolation to larger $r$ is indicated
by a dashed line, Xue et al (2011, 2014), Brown, Geller \& Kenyon (2014), and \cite{deason_veil}, are shown.
The text within the figure indicates the colors used for each reference
and the nature of the sample in each case.
   \label{figure_allsigma} }
\end{figure}

\clearpage

\section{Implications}

One of our goals is to determine the radial distribution of the outer halo
of the Milky Way using RR Lyrae as tracers of its stellar population,
and to use this information to derive the total mass of the MW.
We calculate distances assuming a fixed $M_R$ of +0.6~mag as the
flux-averaged mean over the period, then correct for interstellar
reddening assuming that the variable is so distant that the
full reddening from the maps of Schlegel, Finkbeiner \& Davis (1998) applies.

Fig.~4 shows a histogram of the full sample
of 464 candidate RR Lyr beyond 50~kpc as a function of log($r$), where $r$ is
the heliocentric distance.  
Both axes have logarithmic scales.  The vertical
axis is the number of variables in
bins equally spaced in distance starting at 50~kpc, with 2.8~kpc/bin.
These counts, accumulated over a solid angle on the sky of $\Omega \sim 10,000$~deg$^2$,
represent
$\rho(r) ~ (\Omega / 4\pi) ~ 4 \pi r^2 {\Delta}r$.
We fit a power law $\rho \propto r^{-\gamma}$, over the range of $r$
we cover, i.e. out to just over 100 kpc.  Therefore a linear fit of
the log(counts) vs log($R_{GC}$) will yield a slope that is
$-\gamma + 2$. Our best fit slope is $\gamma = 3.8\pm{0.3}$.

\begin{figure}
\epsscale{0.8}
\plotone{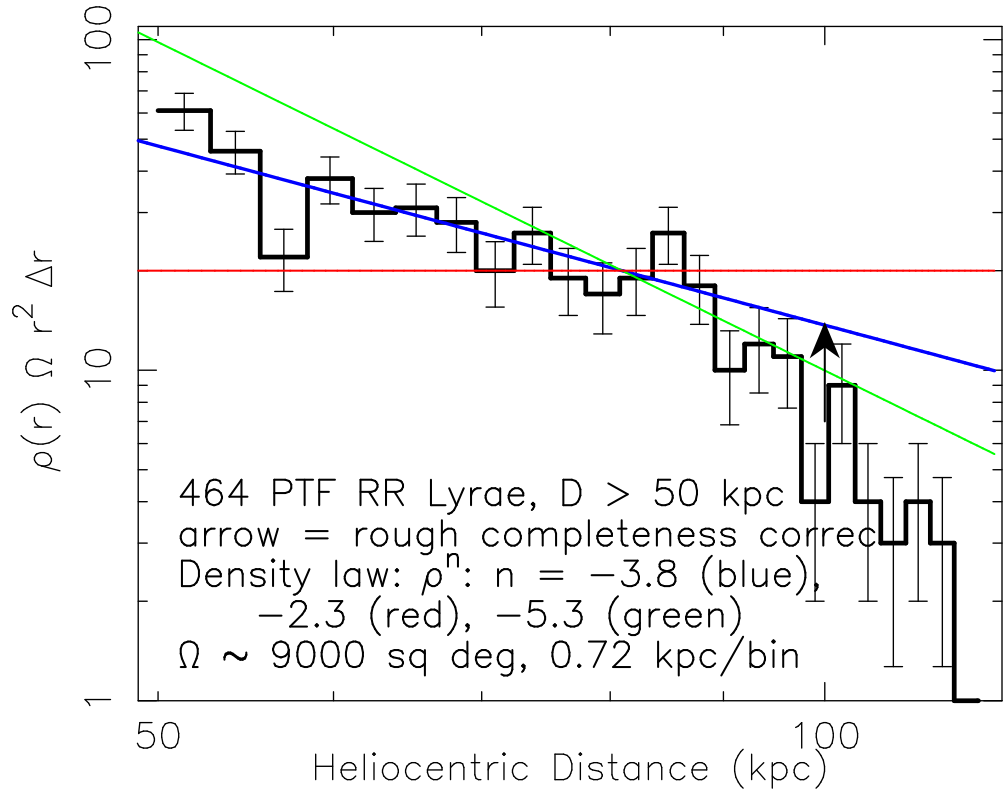}
 \caption{The number density of RR Lyr as a function of distance is shown together
with several power law fits which suggest $\gamma \approx 3.8$.  A lower limit
to the incompleteness correction at a distance of 100~kpc is shown
as a vertical arrow.
   \label{figure_dist_hist} }
\end{figure}

The completeness corrections are very difficult to calculate because the number
of epochs observed varies greatly among the PTF fields, depending on which
(other) project requested observations of the field; the galactic halo project has until recently
been allocated no P48 imaging time at all; we have
simply used observations acquired for other PTF subprojects
and thus have no control over the cadence of observing nor the fields observed.
Furthermore, PTF observations at the P48 are carried out
as long as the dome can safely remain open irrespective of clouds.  Thus
the limiting magnitude of each exposure of a given field has a very wide range,
with clouds (i.e. a bright limiting magnitude) more common than exceptional
nights which are perfectly clear and with good seeing, which have the faintest
limiting magnitudes.  
An RR Lyr at 100~kpc has a mean R of 20.6, close to the limit of the
PTF survey, so that in many PTF images it will not be detected.  To aquire a specified
minimum number $N$  of R detections, where $N$ is set by our search procedure
to identify a candidate RR Lyr, will thus
require many more epochs of observation with PTF of its field 
than would be the case for a RR Lyr at 50~kpc in the same field.
We indicate our best guess of a lower limit to the completeness correction
arising from this issue
at 100~kpc by the vertical arrow in Fig.~4.  Other such issues will further increase
the completeness correction above 90~kpc.

The only effort that reaches out to the radial range covered by our
RR Lyr sample is that of Deason et al (2014), who claim there is a very steep
outer halo profile, with $\gamma \sim 6$ beyond $r = 50$~kpc and even
steeper slopes $\gamma \sim 6-10$ at larger radii.
Their sample of $\sim$5200 stars is contaminated by QSOs; 
a photometric separation is not sufficient and cuts out many BHB stars.

We cannot reproduce the extremely steep decline in $n(r)$ that they claim to observe.
We believe that with our RR Lyr sample selected through variability, the precise distances
we obtain for our RR Lyr stars, and our low QSO contamination, that our result
that $\rho \propto r^{-3.8\pm0.3}$ for $50 < r < 100$~kpc is correct.
Most other recent analyses, i.e. Xue et al (2015), Brown et al 2010,
find $\gamma$ between 3.5 and 4.5 from $r=20$~kpc
out to the limit of their data, between 60 and 80~kpc; see e.g. Fig.~1 of Gnedin et al (2010).

As a rough indication of the total mass of the MW, we assume a spherical halo,
ignoring the subtleties in the inner part of the MW of the thick and thin disk.
We also ignore the difference between our heliocentric distances and galactocentric distances.
The standard way to obtain the total enclosed mass given a set of tracers,
be they globular clusters or RR Lyr or any other low mass objects in
the outer halo of a massive galaxy, is to solve the Jeans equation.
Watkins, Evan \& An (2010) have solved the Jeans equation for the case of 
set of mass tracers with both distances (not projected distances)
and radial velocities
located in the outer halo of a massive galaxy such as the Milky Way.
To accomplish this they assume that the tracers follow
a power-law density distribution $\rho \propto r^{-\gamma}$.
They further assume that  a NFW halo
(Navarro, Frenk \& White 1996)
is an adequate representation of 
the outer part of the Milky Way halo.  

Their  result then simplifies (see Evans, An, \& Deason 2011) to 

$$ M_{vir} \approx { {r_{vir}^{0.5} ~ (0.5 + | \gamma | - 2 \beta}) \over {G N} } 
  ~ {\Large{\sum_{i=1}^{N} r_i^{0.5} v_{r,i}^2} }. ~~~~~~ \rm{(eqtn.~1)} $$

Lacking proper motions, we cannot evaluate
the velocity anisotropy $\beta$; we assume isotropic orbits ($\beta = 0$) for the 
outer halo RR Lyr.  Attempts to measure $\beta$ in the inner halo are more successful
as much larger samples of tracers can be assembled, usually SDSS BHB stars.
Kafle et al (2012), among others, suggest $\beta = +0.5$ from  $\sim$25~kpc
out to the limit of their sample at $\sim$60~kpc, a value similar to that deduced by
Williams \& Evans (2015), who apply a new theoretical analysis to previously published data,
while Deason et al (2011) suggest from
CDM simulations that $\beta = 0$.

Using the above formula with $\beta = 0.0$ we find a total mass out to 110~kpc
of $\sim{7{\pm{1.5}}\times}10^{11}~M_{\odot}$, where the uncertainty corresponds
to a range in $\beta$ from $-0.5$ to +0.5.
This suggests a rather low total mass for the MW, but is in
reasonable agreement with several recent determinations based on studying
halo stars at smaller radii (see the compiled recent measurements in Fig.~5 of
Williams \& Evans 2015), although analyses including the outermost MW satellites and M31 timing
arguements continue to suggest a higher total mass.

Future work should expand our sample of outer halo RR Lyr significantly.
Another two years of PTF imaging (2012--2014) has recently been
added to the PTF variable star database, suggesting it is time to
conduct another search for candidate RR Lyr in the PTF+iPTF database,
which is a rather daunting task given the limited available human resources.

\end{document}